# Free-Space Nonlinear Beam Combining Towards Filamentation

Shermineh Rostami,[1] Wiktor Walasik,[2] Daniel Kepler,[1] Matthieu Baudelet,[1,3]

Natalia M. Litchinitser,[2] and Martin Richardson[1*]

[1] *Laser Plasma Laboratory, Townes Laser Institute, College of Optics and Photonics,*

*University of Central Florida, USA*

[2] *Department of Electrical Engineering, University at Buffalo, The State University of New York, Buffalo, New York 14260, USA*

[3] *National Center for Forensic Science /Chemistry department, University of Central Florida, USA*

*Corresponding author:* mcr@creol.ucf.edu

**Multi-filamentation opens new degrees of freedom for manipulating electromagnetic waves in air. However, without control, multiple filament interactions, including attraction, repulsion or fusion often result in formation of complex disordered filament distributions. Moreover, high power beams conventionally used in multi-filament formation experiments often cause significant surface damage. The growing number of applications for laser filaments requires fine control of their formation and propagation. We demonstrate, experimentally and theoretically, that the attraction and fusion of ultrashort beams with initial powers below the critical value enable the eventual formation of a filament**

**downstream. Filament formation is delayed to a predetermined distance in space, avoiding optical damage to external beam optics while still enabling robust filaments with controllable properties as if formed from a single high power beam. This paradigm introduces new opportunities for filament engineering eliminating the need to use high peak power lasers sources.**

Filament generation in air is a nonlinear process requiring high power ultrashort laser pulses (> 3 GW for 800 nm) [1, 2] to induce a dynamic balance between Kerr self-focusing and plasma defocusing. The unique properties of these self-channeled structures, such as diffraction-free propagation and the formation of plasma channels many times longer than the Rayleigh length, allow for the development of a wide variety of applications [3]. Arrays of filaments have been shown to enable various guided-wave structures in free space for visible [4], infrared [5], radio and microwave [6-8] frequency radiation, depending on their configuration. The generation of large arrays of filaments calls for phase manipulation of the laser pulses as well as precise engineering of intensity/phase distributions and the nonlinear beam interactions [8-14].

Several studies have focused on enhancing the unique filament characteristics by manipulating the initial beam properties such as the spatial intensity and/or phase profile, the wavelength, and the polarization [15-20], which can be extended to the case of multiple filaments. Indeed, the close proximity (comparable to the beam waist) of two or more filaments has been shown to give rise to several new phenomena, such as interference, energy exchange, attraction, repulsion, spiral motion, break up and fusion of multiple filaments [21-24]. The separation between the intensity and/or phase perturbations in the beam (beam size) was shown to have an effect on the growth of multiple filaments, diameter of the plasma channel and overall electronic density [25]. Phase controlled interactions between crossed and parallel filaments in Kerr media have been studied theoretically and experimentally [23, 26-33]. Enhanced THz generation was observed by controlling spatial and temporal phase between two filaments extended to larger arrays [34]. White light generation was controlled with the relative delay and polarization of two crossed femtosecond laser beams in flint glass [29]. The fusion of two in-phase filaments was shown to even prolong the propagating distance and enhance the channel stability in comparison to a single filament [23, 33].

Despite these extensive studies performed with single and multiple filaments, controlling their formation and the inevitable optical damages induced by the filamenting beams remain the most significant fundamental and technological challenges. We propose a solution to these issues by exploiting nonlinear coherent combining of beams with initial powers below the critical value, $P_{\text{cr}}$, to create filaments. This is the first experimental demonstration of attraction and fusion of ultrashort beams towards robust formation of a single filament. This approach opens up new opportunities to induce filamentation by combining multiple low power beams having more control over the filament and the associated by-products (e. g. plasma channel, spectrum) with the practical advantage of avoiding the optical damage of beam-steering optical components usually encountered with high power beams.

The experiment was conducted using the Multi-Terawatt Femtosecond Laser (MTFL) [35] at the University of Central Florida, with an output of 800 nm, 45 fs (full width at half-maximum—FWHM) pulses at 10 Hz. The horizontally polarized laser pulses were rotated with a half-wave plate ($\lambda/2$) to allow for maximum transmission through the interferometer (Fig. 1). A beam splitter (BS) was then used to separate the beam between two arms. After travelling equal distances and passing through identical focusing lenses ($f = 5$ m), the two beams were recombined at the second beam splitter. The latter and one of the lenses were mounted on a translation stage to enable fine control of the spatial separation of the centroids of the two beams, keeping the geometrical focus fixed. The relative phase between the two arms was tuned with controllable spatial and temporal separations between the beams to a resolution of 25.4 μm and ~7 fs, respectively. The two linearly polarized Gaussian beams each with 4.25 mm diameter (FWHM) in the transverse plane, and starting with $0.72 P_{cr}$ and $0.68 P_{cr}$ peak powers, were spatially brought into horizontal proximity to one another in steps of 25.4 μm and overlapped in time for each step. To reduce spatial positioning errors, one of the arms (beam S with $0.72 P_{cr}$) was kept fixed and only the other arm (beam D with $0.68 P_{cr}$) was moved relative to beam S (Fig. 2A).

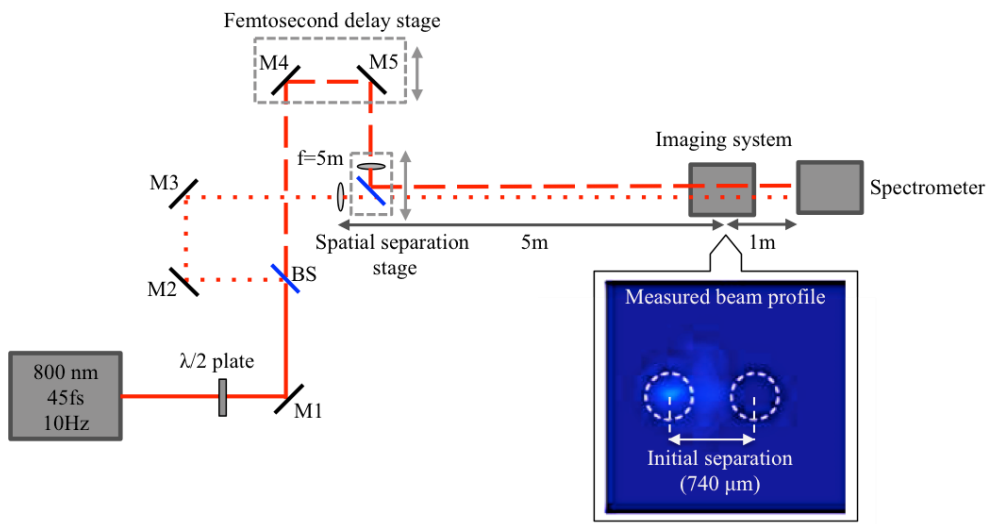

*Figure 1. Experimental setup. The inset shows the beam profile measured at the geometrical focus when the initial separation between the two beams was 740μm.*

The overall beam profile (Fig. 2A) was measured at the geometrical focus, using a series of grazing incidence uncoated fused silica wedges as an attenuator. The large angle of incidence relative to the surface normal prevents surface damage from the high intensity filaments. The low reflectivity attenuates the beam intensity by a factor of $10^7$ allowing the CCD to safely acquire the filament beam profile with a lens ($f = 100$ mm) [36].

Figure 2 presents the experimental and simulated (see Methods) normalized intensity profile at the geometrical focus, for various initial separation distances between the beams. For the initial separation larger than 1250 μm, we observe negligible deviations from the initial separation (dashed lines in Figs. 2A&B) after 5 meters of propagation indicating that the interaction between the beams is weak for large separation distances. Between 1250 and 330 μm of initial separation, the attraction between the beams becomes stronger as their separation at the focal point is smaller than the initial separation. The small discrepancy between the simulation and experimental results is due to the fact that, in the experimental case, the final intensity distribution at the focusing plane was statistically calculated as a result of 500 laser shots. On the contrary, in the simulations only one laser shot (without noise) was simulated. Below 330 μm of initial beam separation, the attraction between the two beams results in their fusion into a stable single filament.

Figure 2C shows the measured (top row) and simulated (bottom row) transverse profiles at selected initial separations. As the distance between the beams decreases from 1500 μm to 660 μm, their separation reduces after 5 meters of propagation until they fuse into a single entity surrounded by a higher energy reservoir that lacks rotational symmetry in the transverse $x$–$y$ plane (see Fig. 2C at 660 μm separation).

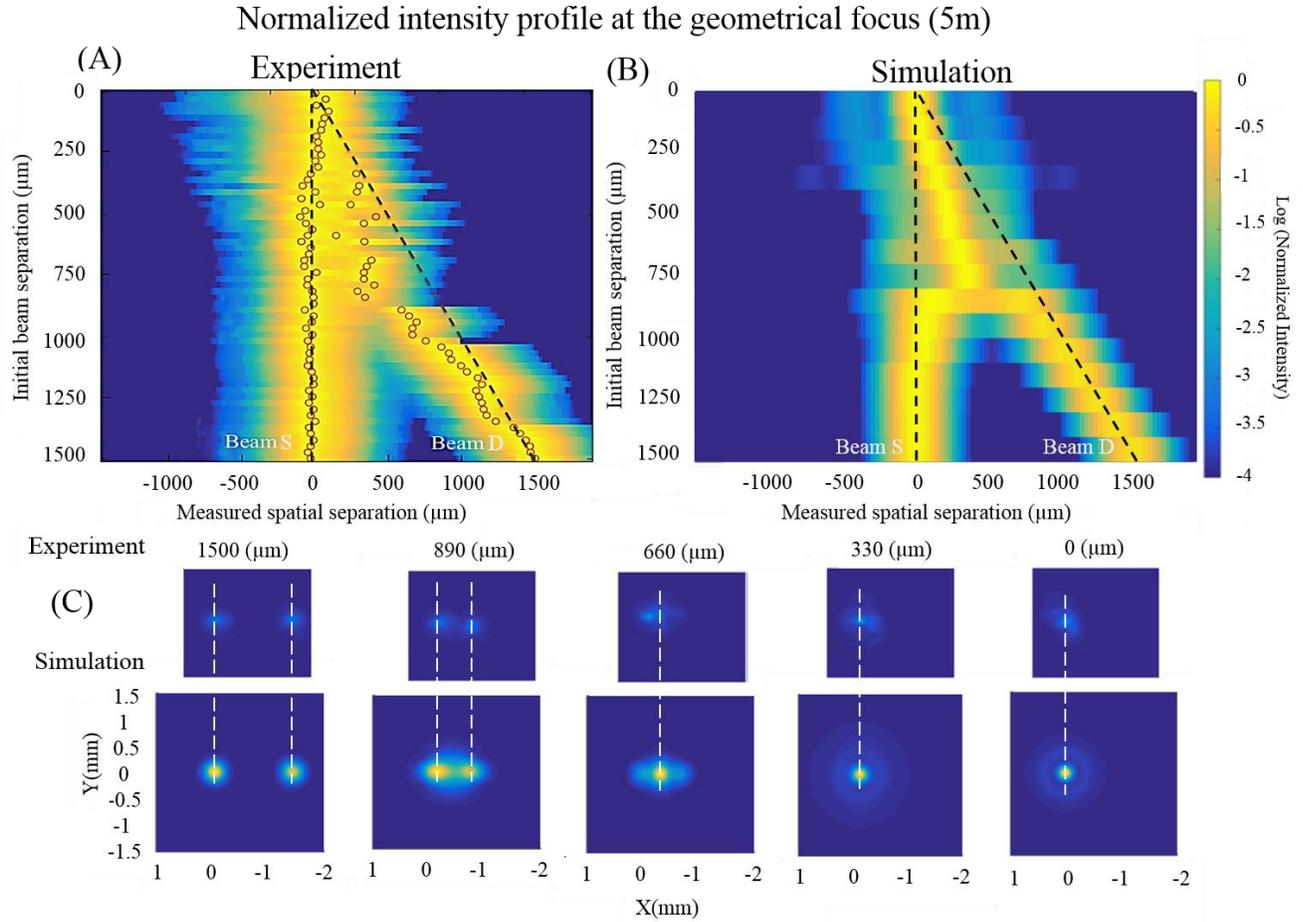

*Figure 2. (A) Measured and (B) calculated normalized intensity profiles of the 2 beams in the transverse direction x at the geometrical focus (z=5 m). Black dashed lines show the initial separation between the stationary (beam S) and dynamic (beam D) beams and red circles denote the local intensity maxima in the experimental profiles. (C) Experimental (top row) and simulated (bottom row) beam profiles in transverse plane (x-y) for 1500, 890, 660, 330 and 0 μm initial separation. White dashed lines indicate the position of the local intensity maxima in the simulation results.*

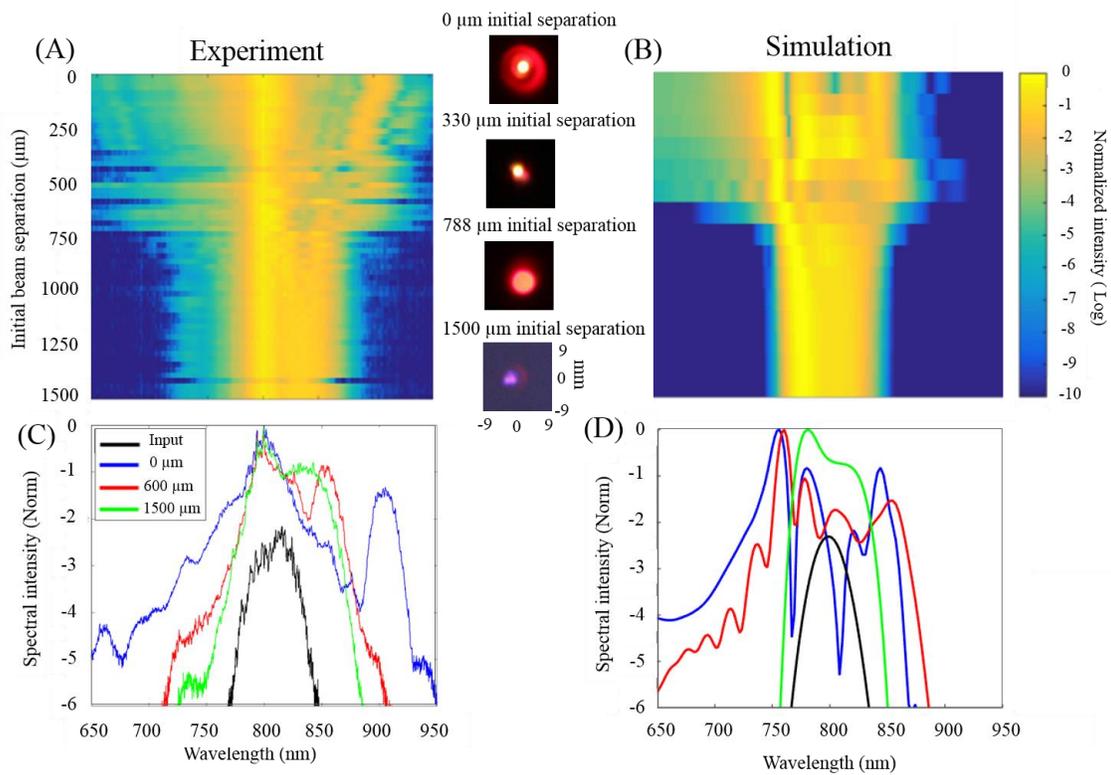

*Figure 3. (A) Measured and (B) calculated spectral intensity of combined beams as a function of initial beam separation. Pictures of the measured conical emission at selected initial beam separation are shown in the middle. (C) Measured and (D) calculated spectrum for selected initial beam separations compared to the spectrum of the individual sub-critical beams.*

In the region of stable single filament generation (330 μm–0 μm), the reservoir has a weaker local intensity but its cross-section is larger with a rotationally symmetric profile in the transverse $x$–$y$ plane (see Fig. 2C at 330 and 0 μm separations).

Figure 3 shows the spectra measured one meter after the geometrical focus. In the simulations, the intensity of the output spectra are averaged over the whole transverse plane ($x - y$) at $z = 6$ m (where the overall spectra were measured experimentally).

Spectral broadening with respect to the input spectrum (black in Figs 3C&D) is observed even in the regime of negligible attraction (green in Figs 3C&D). In this regime, the spectral broadening due to the co-propagation of the two beams is negligible when compared to the propagation of a single beam with the same overall energy (see Supplementary Fig.1). Starting in the region of unstable attraction (beams initially closer than 850 µm), the spectrum broadens both towards the visible and the infrared. This broad supercontinuum is an indication of the existence of nonlinear effects not seen during propagation of the individual, sub-critical beams.

The transition between regimes of attraction and fusion was theoretically shown by Bergé *et al.* [30, 31]. Using their nomenclature and considering two in-phase beams, a critical separation value, $\delta_c$, below which the fusion occurs, can be derived from their work. In our experiments the beams had a radius of $\rho = 0.2$ mm (measured in the focal plane at $z = 5$ m) and $P_1 = 0.72\, P_{cr}$ and $P_2 = 0.68\, P_{cr}$ (corresponding to $N_1 = 2.88\pi$ and $N_2 = 2.72\pi$). For these parameters, the critical fusion distance $\delta_c$ can be calculated as a root of Eq. (14) in Ref. [30] and is equal to $\delta_c \approx 600$ µm. It can be seen from Fig. 3 that the sudden broadening of the spectrum starts at approximately 700 µm and peaks at around 600 µm, confirming the presence of nonlinearities induced by the beam combination. These observations also present the ability now of controlling the spectral broadening with the separation between the two beams while keeping their initial intensity constant and below the critical value for filamentation.

Figure 4 shows a comparison of the maximum light intensity as a function of the propagation distance z (Figs. 4A&C) and the normalized spectral intensity at $z = 6$ m (Figs. 4B&D) for a single filament created from a single beam (blue) and a filament created by nonlinear combination of two subcritical beams (green). The input energy of the single beam is chosen based on the resulting energy of the two interfering input Gaussian beams. As shown earlier, for initial separations larger than 700 μm, the two beams do not fuse, preventing the generation of a filament. As the initial separation decreases below 700 μm the two beams start to fuse and create a single filament. For instance, with a beam separation of 600 μm (shown in Figs. 4A&B), the total input energy is 0.52 mJ and the filament is created. The maximum light intensity of a filament created by combining the two beams is lower than the intensity of a filament created using a single beam with 0.5 mJ energy, but the filament resulting from the fusion starts earlier, is longer (length measured as the distance over which the light intensity is larger than half of the maximum value) and results in a similar supercontinuum spectrum 1 m after the geometrical focus (Fig. 4B). For smaller initial separations (e.g. 400 µm shown in Figs. 4C&D), the discrepancy between the two filaments is negligible, either they were created from a single pulse of 0.65 mJ or two pulses of 0.32 mJ. The maximum intensity and the length of the filaments are very similar with comparable spectra, even if the filament resulting from the fusion of the two beams is created a little earlier.

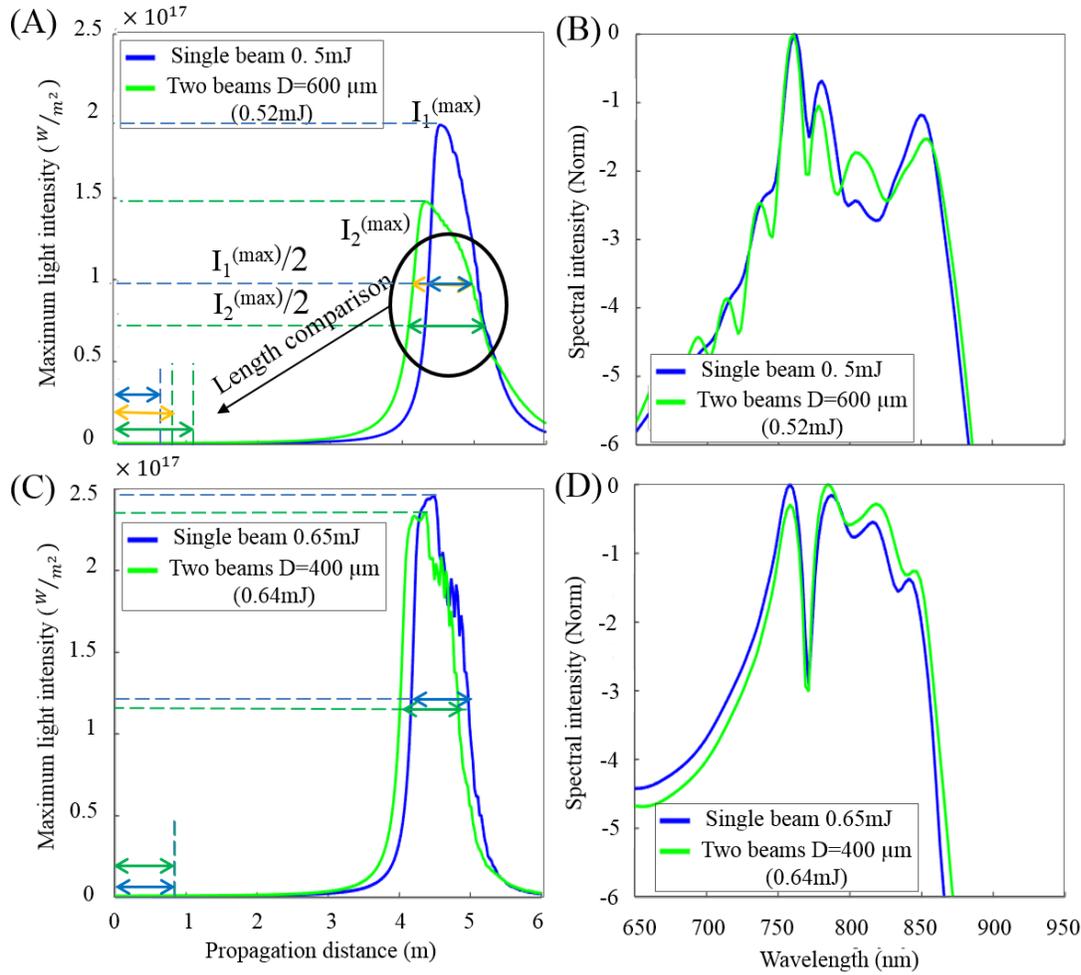

*Figure 4. (A) and (C) calculated intensity along propagation for a single beam (blue) and two combined beams (green) with comparable powers. (B) and (D) spectral intensity for a single beam (blue) and two combined beams (green) with comparable powers.*

In summary, we showed the significance of using sub-critical pulses for the creation and control of filaments: pulses with total energy equal to that of a single beam but distributed over multiple beams separated by a precisely controlled distance are shown to induce identical spectral broadening. Furthermore, the starting point and the length of the filament created from the fusion of the two lower powered beams can be controlled with the initial separation between their centroids. This new paradigm in filament formation opens up new opportunities for precise engineering of filamentation using low power laser pulses.

**Methods**

The numerical simulations were performed using a (3+1) dimensional split-step Fourier scheme to solve the Nonlinear Schrödinger Equation describing the evolution of the slowly varying envelope of the electric field $E(x, y, z, t)$ of ultrashort pulses propagating in air [1]:

$$\frac{\partial E}{\partial z} = \frac{i}{2k_0}\left(\frac{\partial^2}{\partial x^2} + \frac{\partial^2}{\partial y^2}\right)E - i\frac{k''}{2}\frac{\partial^2 E}{\partial t^2} + ik_0 n_2 \mathcal{R}(t)E - \left(\frac{\sigma_B}{2} + i\frac{k_0}{2\rho_c}\right)\rho E - \frac{\beta^{(K)}}{2}|E|^{2K-2}E.$$

Here, $E$ is normalized in such a way that $|E|^2$ represents the light intensity expressed in W/m$^2$. The total electric field is given as $\mathcal{E}(x,y,z,t) = E(x,y,z,t)\exp(-i\omega_0 t)$, where $\omega_0 = \frac{2\pi c}{\lambda_0}$ denotes the light angular frequency, $c$ is the speed of light in vacuum, and $\lambda_0 = 800$ nm is the central free-space wavelength of the femtosecond pulse. $k_0 = \frac{2\pi}{\lambda_0}$ is the free space wavenumber, $k'' = 2$ fs$^2$/m [37] and $n_2 = 5.57 \times 10^{-19}$ cm$^2$/W [38] are the group velocity dispersion and nonlinear index of refraction for air at the wavelength $\lambda_0$, respectively. The inverse Bremsstrahlung cross-section responsible for plasma absorption is denoted by $\sigma_B = 5.47 \times 10^{-20}$ cm$^2$, the critical plasma density at which an air plasma becomes transparent is given by $\rho_c = 1.74 \times 10^{21}$ cm$^{-3}$, and the multiphoton absorption coefficient $\beta^{(K)} = 3.67 \times 10^{-95}$ cm$^{13}$/W$^7$, where $K = 8$ shows the number of photons needed to be absorbed to overcome the effective ionization potential of air $U_i = 11$ eV [37]. The nonlinear response of air is described by

$$\mathcal{R}(t) = (1-\alpha)|E(t)|^2 + \frac{\alpha\,(\Gamma^2 + \omega_R^2)}{\omega_R} \int_{-\infty}^{t} \exp[-\Gamma(t-\tau)]\sin[\omega_R(t-\tau)]\,|E(\tau)|^2\,d\tau,$$

taking into account both the instantaneous Kerr effect and the delayed response due to stimulated molecular Raman scattering with the weights $(1-\alpha)$ and $\alpha = 0.5$, respectively. The characteristic relaxation time for oxygen molecules is taken as $\Gamma^{-1} = 70$ fs and the molecular response frequency is given by $\omega_R = 16$ THz. The time evolution of the electron density is governed by equation

$$\frac{\partial \rho}{\partial t} = \sigma_K \rho_{\text{nt}} |E(t)|^2 + \frac{\sigma_B}{U_i} \rho |E(t)|^2,$$

where $\sigma_K = \beta^{(K)}/(K\hbar\omega_0 \rho_{\text{nt}})$ is the multiphoton ionization coefficient, $\hbar\omega_0$ denotes the energy of a single photon at the frequency $\omega_0$, and $\rho_{\text{nt}} = 5 \times 10^{18}$ cm$^{-3}$ is the density of neutral oxygen molecules. The numerical values of the parameters for air were taken from Ref. [1]. Using the values presented here, starting with a 45 fs Gaussian beam (4.25 mm FWHM) focused by a 5 m lens in air, the critical peak power for filamentation was found to be $P_{cr} = 6.3$ GW.

**ACKOWLEDGMENTS**

This work was funded by the U.S. Army Research Office (MURI W911NF12R001202 and W911NF110297), the HEL-JTO (MRI FA05501110001) and the State of Florida.

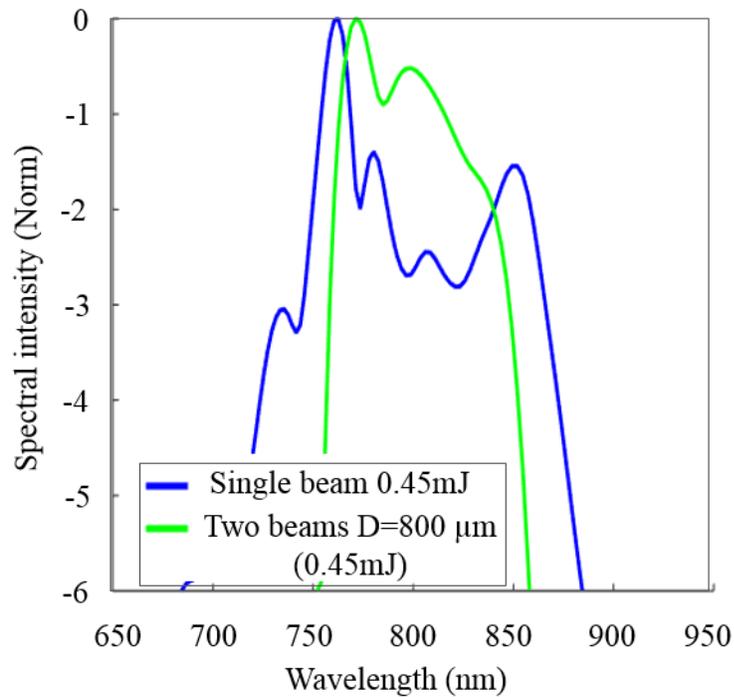

*Supplementary Fig.1: Spectral intensity for a single beam (blue) and two combined beams (green) with comparable powers with initial separation of 800 μm.*

REFERENCES

[1] Couairon A. and Mysyrowicz, A. Femtosecond filamentation in transparent media. *Phys. Rep,* **441** (2-4), 47 (2007).

[2] Liu, W. & Chin, S. L. Direct measurement of the critical power of femtosecond Ti:sapphire laser pulse in air. *Opt. Express* **13** (15), 5750, (2005).

[3] Chin, S. L. *et al.* Advances in intense femtosecond laser filamentation in air. *Laser Phys.* **22**, (1), 1, (2011).

[4] Jhajj, N., Rosenthal, E. W., Birnbaum, R., Wahlstrand, J. K. & Milchberg, H. M. Demonstration of long-lived high-power optical waveguides in air. *Phys. Rev. X* **4**, 011027, (2014).

[5] Alshershby, M., Hao, Z., Camino, A. & Lin, J. Modeling a femtosecond filament array waveguide for guiding pulsed infrared laser radiation. *Opt. Commun.*, **296**, 87 (2013).

[6] Alshershby, M., Hao, Z. & Lin, J. Guiding microwave radiation using laser-induced filaments: the hollow conducting waveguide concept. *J. Phys. D: Appl. Phys*. **45**, 265401 (2012).

[7] Ren, Y., Alshershby, M., Hao, Z., Zhao, Z. & Lin, J. Microwave guiding along double femtosecond filaments in air. *Phys. Rev. E* **88**, 013104 (2013).

[8] Kudyshev, Z. A., Richardson, M. C. & Litchinitser, N. M. Virtual hyperbolic metamaterials for manipulating radar signals in air. *Nat. Commun*. **4**, 2557 (2013).

[9] Camino, A., Xi, T., Hao, Z. & Lin, J. Femtosecond filament array generated in air. *Appl. Phys. B* **121**, 363 (2015).

[10] Apeksimov, D. V. *et al.* Control of the domain of multiple filamentation of terawatt laser pulses along a hundred-meter air path. *Quantum Elec*. **45** (5), 408 (2015).

[11] Fu, Y. *et al.* Control of filament branching in air by astigmatically focused femtosecond laser pulses. *Appl. Phys. B* **103** (2), 435 (2011).

[12] Grow, T. D. and Gaeta, A. L. Dependence of multiple filamentation on beam ellipticity. *Opt. Express* **13**, 4594 (2005).

[13] Barbieri, N. *et al.* Double helical laser beams based on interfering first-order Bessel beams. *J. Opt. Soc. Am. A* **28**, 1462, (2011).

[14] Walasik, W. and Litchinitser, N. M. Dynamics of Large Femtosecond Filament Arrays: Possibilities, Limitations, and Trade-Off. *ACS Photonics* **3**, 640 (2016).

[15] Kaya, G. *et al.* Extension of filament propagation in water with Bessel-Gaussian beams. *AIP Advances* **6**, 035001 (2016).

[16] Polynkin, P., Kolesik, M., Moloney, J., Siviloglou, G. & Christodoulides, D. Curved Plasma Channel Generation Using Ultraintense Airy Beams. *Science* **324**, 229 (2009).

[17] Polynkin, P., Kolesik, M. & Moloney, J. Filamentation of Femtosecond Laser Airy Beams in Water. *Phys. Rev. Lett.* **103**, 123902 (2009).

[18] Polynkin, P. *et al.* Generation of extended plasma channels in air using femtosecond Bessel beams. *Opt. Express* **16**, 15733 (2008).

[19] Rostami, S. *et al.* Dramatic enhancement of supercontinuum generation in elliptically-polarized laser filaments. *Sci. Rep.* **6**, 20363 (2016).

[20] Panagiotopoulos, P., Whalen, P., Kolesik, M. & Moloney, J. Super high power mid-infrared femtosecond light bullet. *Nat. photon.* **9**, 543 (2015).

[21] Tzortzakis, S. *et al.* Breakup and Fusion of Self-Guided Femtosecond Light Pulses in Air. *Phys. Rev. Lett.* **86** (24), 5470 (2001).

[22] Hosseini, S. A. *et al.* Competition of multiple filaments during the propagation of intense femtosecond laser pulses. *Phys. Rev. A* **70**, 033802 (2004).

[23] Xi, T.-T., Lu, X. & Zhang, J. Interaction of Light Filaments Generated by Femtosecond Laser Pulses in Air. *Phys. Rev. Lett.* **96**, 025003 (2006).

[24] Georgieva, D. A. and Kovachev, L. M. Energy transfer between two filaments and degenerate four-photon parametric processes. *Laser Phys*. **25** (3), 035402 (2015).

[25] Kosareva, O. G. *et al*. Controlling a bunch of multiple filaments by means of a beam diameter. *Appl. Phys. B* **82**, 111 (2006).

[26] Ishaaya, A. A., Grow, T. D., Ghosh, S., Vuong, L. T. & Gaeta, A. L. Self-focusing dynamics of coupled optical beams. *Phys. Rev. A* **75**, 023813 (2007), and references therein.

[27] Shim, B. *et al*. Controlled interactions of femtosecond light filaments in air. *Phys. Rev. A*, **81**, 061803(R) (2010).

[28] Barbieri, N. *et al*. Helical Filaments. *Appl. Phys. Lett*., **104** (26), 261109 (2014).

[29] Kolomenskii, A. A. *et al*. White-light generation control with crossing beams of femtosecond laser pulses. *Opt. Express* **24** (1), 282 (2016).

[30] Bergé, L., Schmidt, M. R., Juul Rasmussen, J., Christiansen, P. L. & Rasmussen, K. O. Amalgamation of interacting light beamlets in Kerr-type media. *J. Opt. Soc. Am. B* **14**, 2550 (1997).

[31] Bergé, L. Coalescence and instability of copropagating nonlinear waves. *Phys. Rev. E* **58** (5), 6606 (1998).

[32] Cai, H. *et al*. Attraction and repulsion of parallel femtosecond filaments in air. *Phys. Rev. A* **80**, 051802(R) (2009).

[33] Ma, Y.-Y., Lu, X., Xi, T.-T., Gong, Q.-H. & Zhang, J. Filamentation of interacting femtosecond laser pulses in air. *Appl. Phys. B* **93**, 463 (2008).

[34] Mitryukovskiy, S. I., Liu, Y., Prade, B., Houard, A. & Mysyrowicz, A. Coherent synthesis of terahertz radiation from femtosecond laser filaments in air. *Appl. Phys. Lett*. **102**, 221107 (2013).

[35] Webb, B. *et al*. Compact 10 TW laser to generate multi-filament arrays. *CLEO*: 2014, OSA Technical Digest (online) (Optical Society of America, 2014), paper SM1F.6.

[36] Lim, K. and Richardson, M. C. Laser filamentation – beyond self-focusing and plasma defocusing. PhD Dissertation, CREOL, The College of Optics and Photonics, Orlando, 2014.

[37] Mlejnek, M., Wright, E. M. & Moloney, J. Dynamic spatial replenishment of femtosecond pulses propagating in air. *Opt. Lett,* **23** (5) (1998).

[38] R. Boyd, Nonlinear Optics 3rd Ed. (Elsevier, 2007).